# Dynamic versus Anderson wavepacket localization

Olga V. Borovkova,[1] Valery E. Lobanov,[1] Yaroslav V. Kartashov,[1,2] Victor A. Vysloukh,[3] and Lluis Torner[1]

[1]*ICFO-Institut de Ciencies Fotoniques, and Universitat Politecnica de Catalunya, Mediterranean Technology Park, 08860 Castelldefels (Barcelona), Spain*
[2]*Institute of Spectroscopy, Russian Academy of Sciences, Troitsk, Moscow Region, 142190, Russia*
[3]*Departamento de Fisica y Matematicas, Universidad de las Americas—Puebla, 72820 Puebla, Mexico*

We address the interplay between two fundamentally different wavepacket localization mechanisms, namely resonant dynamic localization due to collapse of quasi-energy bands in periodic media and disorder-induced Anderson localization. Specifically, we consider light propagation in periodically curved waveguide arrays on-resonance and off-resonance, and show that inclusion of disorder leads to a gradual transition from dynamic localization to Anderson localization, which eventually is found to strongly dominate. While in the absence of disorder, the degree of localization depends critically on the bending amplitude of the waveguide array, when the Anderson regime takes over the impact of resonant effects becomes negligible.

## 1. Introduction

External potentials, especially periodic ones, generate a variety of unique phenomena associated with the transport and evolution dynamics of localized wavepackets. Of particular interest is the evolution of wave excitations of different physical nature propagating in inhomogeneous media, including light waves in materials with inhomogeneous refractive index and matter waves in Bose-Einstein condensates held in optical or magnetic traps. Recent progress in the fabrication of periodic optical materials, such as shallow photonic lattices and photonic crystals with high refractive index contrast, has enormously enriched the possibilities to study the propagation of light waves, including their diffraction, transmission, reflection and confinement [1-3]. Importantly, the analogies existing between the equations governing light propagation in periodic refractive index landscapes, Bose-Einstein condensates in external potentials, and electrons in crystalline lattices afford the exploration and observation of a number of previously elusive phenomena relevant to other areas of physics [4].

One of such phenomena is the *dynamic wavepacket localization* initially introduced for electrons moving in conducting crystals under the influence of an external ac electric field [5,6]. The effect was later observed with confined light beams propagating in curved waveguide arrays, where the periodic curvature variation emulates the action of the ac electric field on electrons [7-11]. The effect is *resonant*, in the sense that it occurs only around specific bending amplitudes and bending periods, which yield a collapse of the quasi-energy band of the dynamical array [8,10]. Under on-resonance conditions wavepackets experience periodic expansion and shrinkage, so that they restore the input profile after each bending period, while off-resonance propagation features slow wavepacket expansion. Illustrating the analogy mentioned above, the same phenomenon occurs in Bose-Einstein condensates in periodically shaken optical lattices [12].

A fundamentally different localization mechanism, termed *Anderson localization*, was discovered in solid-state physics for electrons in time-independent disordered potentials [13,14]. This is universal phenomenon that may occur for waves of different physical nature, since it is caused by the localization of eigenmodes of the disordered potential in one-and two-dimensional settings [15]. Anderson localization has been observed with microwaves [16,17], matter [18,19] and acoustic [20] waves, and in disordered optical lattices [21–30]. A particularly important manifestation of this effect is the *transverse* Anderson localization [22] that can be achieved in optical lattices disordered only in the transverse direction, but not in the direction of light propagation, similarly to the time-independent potentials considered in [13,14]. Transverse Anderson localization has been reported in optically induced lattices [24], fabricated [25] and laser-written [26] waveguide arrays, and in plasmonic nanowires [29], to name just a few settings. Inclusion of transverse disorder suppresses Bloch oscillations in lattices with transverse gradients [31] and even prevents resonant delocalization in such structures if the coupling constants are longitudinally modulated [32].

Here we are interested in the interplay between the two mentioned mechanisms. On physical grounds, the presence of disorder in a dynamically-varying refractive index landscape implies that at each propagation step the wavepacket has to adapt to a new refractive index profile. Therefore, disorder may impact dynamical localization differently on-resonance and off-resonance, or longitudinal variations of the disordered potential may be expected to have a detrimental effect on disorder-induced localization, especially when such variations occur on a scale much shorter than the distance at which the onset of Anderson localization occurs. For example, rapidly varying disorder may result in enhanced expansion (transport) of the input wavepacket, instead of Anderson localization [33]. Therefore, a question arises about the interplay and gradual transition between the two different localization mechanisms and, specifically, to which extent resonant localization, caused by the dynamically varying refractive index landscape, is affected by different types and strength of disorder.

In this paper we report the results of comprehensive series of numerical experiments conducted to address such interplay, and we expose the gradual transition from dynamic to Anderson localization that occurs with both, diagonal or off-diagonal disorder. Disorder significantly enhances localization for bending amplitudes detuned from resonance. On-resonance, inclusion of small disorder enhances delocalization slightly, which however is replaced by localization due to the Anderson mechanism at larger disorder levels. For low and moderate disorders, the degree of localization at a given disorder level is a nonmonotonic function of the bending amplitude, but disorder-induced localization eventually dominates. Our results are relevant to several physical settings, including Bose-Einstein condensates held in shaken optical lattices, and electronic systems in external a.c. fields, where the above-mentioned localization mechanisms can act simultaneously.

## 2. Theoretical model

Our analysis is based on the Schrödinger equation for the dimensionless amplitude $q$ of a wavepacket propagating along the $\xi$-axis in a disordered curved waveguide array or a Bose-Einstein condensate held in suitable optical lattices, which writes:

$$i\frac{\partial q}{\partial \xi} = -\frac{1}{2}\frac{\partial^2 q}{\partial \eta^2} - R(\eta,\xi)q. \tag{1}$$

For concreteness, here we focus on the optical realization in waveguide lattices with a shallow refractive index modulation [1-3], in which case the transverse coordinate $\eta$ and propagation distance $\xi$ are normalized to the characteristic transverse scale $x_0$ and diffraction length $L_{\text{dif}} = kx_0^2$, respectively, with $k = 2\pi n/\lambda$ being the wavenumber and $n$ being the unperturbed refractive index. The dynamic array $R(\eta,\xi) = \sum_{m=-\infty}^{+\infty} p_m G[\eta - \eta_m(\xi)]$ is composed of a set of curved super-Gaussian waveguides $G(\eta) = \exp(-\eta^6/w_\eta^6)$ with width $w_\eta$, waveguide depths $p_m \sim \delta n_m k^2 x_0^2 / n$, where $\delta n_m$ is the actual refractive index modulation depth in the $m$-th waveguide, and waveguide center positions $\eta_m(\xi) = \eta_m(0) + A[\cos(2\pi\xi/T) - 1]$, where $A$ is the bending amplitude and $T$ is the bending period. We consider two different types of disorder, termed diagonal and off-diagonal. In the case of diagonal disorder, the refractive indices $p_m = p(1+\delta p_m)$ of individual waveguides are randomized; in this expression $p$ is the mean refractive index value, while the random variable $\delta p_m$ is uniformly distributed within the segment $[-p_d; +p_d]$. For off-diagonal disorder the coordinates of the waveguide centers $\eta_m(0) = d(m+\delta\eta_m)$ are randomized; here $d$ is the array period and the random relative shift of the waveguide position $\delta\eta_m$ is uniformly distributed within the segment $[-s_d; +s_d]$. The strengths of the diagonal and off-diagonal disorder are controlled by the parameters $p_d$ and $s_d$, respectively. While the entire refractive index landscape in Eq. (1) varies with propagation distance $\xi$, it is important to note that it remains frozen in the coordinate frame oscillating together with the array. Here we set $p_0 = 7.5$, $w_\eta = 0.5$, $d = 2.0$, $T = 60$, which for the characteristic scale $x_0 = 10$ $\mu$m corresponds to an actual refractive index contrast of $\sim 9\times 10^{-4}$, a waveguide width of $\sim 5$ $\mu$m and a lattice period of $\sim 20$ $\mu$m. A normalized propagation distance $\xi = 1$ at the wavelength $\lambda = 632$ nm corresponds to the actual length $\sim 1.44$ mm ($T \sim 86.4$ mm). For the sake of generality and wide applicability, the parameters used in our calculations have been chosen to stand for a general, standard configuration. Also, shallow waveguide arrays with the studied characteristic can be readily fabricated using widely accessible direct laser-writing techniques. In particular, fabrication of curved waveguides in silica glass was reported in [11] and introduction of disorder in such a system is achieved by changing randomly the writing velocity (see [31] for details).

## 3. Results and discussion

Dynamic localization is resonant effect caused by the collapse of the quasi-energy band of the longitudinally modulated waveguide array. Using the tight-binding approximation for a regular array, one can simplify the continuous Eq. (1) to $idc_m/d\xi = -\kappa(c_{m+1}+c_{m-1}) + g(\xi)mc_m$ [10], where $c_m$ are the field amplitudes in the different waveguides, $\kappa$ is the coupling constant, and $g(\xi) = g_0 \cos(2\pi\xi/T)$ is a function determining bending low of the array [here $g_0 \sim A$]. The substitution $c_m(\xi) = \nu_m(\xi)\exp[-im\int_0^\xi g(z)dz]$ transforms the above mentioned tight-binding model into $id\nu_m/d\xi = -(\mathcal{Q}^*\nu_{m+1} + \mathcal{Q}\nu_{m-1})$, where the effective coupling constant $\mathcal{Q} = \kappa \exp\int_0^\xi ig(z)dz$ is introduced. Since $\mathcal{Q}(\xi)$ is a periodic function of $\xi$, the solutions are given by the Floquet-Bloch modes $\nu_m = u_m(\xi)\exp(ib\xi + ik_tmd)$ with dispersion relation $b(k_t) = (2/T)\text{Re}[\exp(-ik_td)\int_0^T \mathcal{Q}(\xi)d\xi]$, where $k_t$ is the transverse wavenumber. This relation can be rewritten in the form $b = 2\kappa J_0(g_0T/2\pi)\cos(k_td)$. The band collapse ($\partial b/\partial k_t \equiv 0$) leading to dynamic localization in real space occurs for the modulation amplitudes at which $J_0(g_0T/2\pi) = 0$ [10]. On physical grounds, such condition corresponds to a vanishing averaged coupling constant over one bending period. Illustrative propagation dynamics in a regular array for a single-site excitation obtained by solving Eq. (1) is shown in Fig. 1(a) for the off-resonant bending amplitude $A = 9.2 < A_r$ and in Fig. 2(a) for the bending amplitude $A = 11.48 = A_r$ that corresponds to the lowest resonance. One can see that in resonance the beam periodically expands across several waveguides, but exactly restores its shape after each oscillation period [Fig. 2(a)]. When the bending amplitude is detuned from resonance the beam expands [Fig. 1(a)]. The width of the beam at $\xi = 16T$ as a function of the bending amplitude is shown in Fig. 3(a), which highlights the resonant nature of the phenomenon.

To elucidate the impact of disorder we perform statistical analysis by generating $Q \sim 500$ realizations of disordered arrays with either diagonal or off-diagonal disorder. In each realization Eq. (1) was integrated for single-site excitations up to considerable distance $\xi = 16T$. We calculated the statistically averaged intensity distribution $I_{\text{av}}(\eta,\xi)$ and the integral form-factor $\chi_{\text{av}}(\xi)$ for the entire propagation path in accordance with the expressions:

$$I_{\text{av}}(\eta,\xi) = Q^{-1} \sum_{i=1}^{Q} |q_i(\eta,\xi)|^2,$$
$$\chi_{\text{av}}(\xi) = Q^{-1} U^{-2} \sum_{i=1}^{Q} \int_{-\infty}^{\infty} |q_i(\eta,\xi)|^4 \, d\eta, \qquad (2)$$

where $U = \int_{-\infty}^{\infty} |q_i|^2 \, d\eta$ is the energy flow that is a conserved quantity of Eq. (1). The inverse form-factor characterizes the width $W_\chi = 1/\chi_{\text{av}}(\xi)$ of the localized core of the averaged intensity distribution, almost disregarding the low-intensity background.

The impact of disorder is pronounced for off-resonant bending amplitudes, as shown in Fig. 1. Although the entire refractive index landscape varies upon propagation, we found that in this setting even weak off-diagonal disorder delays beam expansion. With increase of $s_d$ two diffracting sidelobes gradually vanish and one observes the appearance of a localized central fraction in the averaged intensity distribution [compare Figs. 1(a) and 1(b)]. Notice that this Anderson-localized fraction of the beam performs specific oscillations in the transverse plane following the axis of the curved disordered array. The width of this localized distribution rapidly shrinks with $s_d$ [Figs. 1(c) and 1(d)]. On physical grounds, note that in this regime the effective coupling constant is large due to off-resonant conditions, therefore the presence of disorder that creates local refractive index defects that support localized modes (whose width decreases with increase of $s_d$) tends to counteract fast diffractive broadening. While in a regular array the width of the beam strongly oscillates even at resonance, attaining local maxima at $\xi = (n+1/2)T$ and minima at $\xi = nT$, in the presence of strong disorder the width of statistically averaged intensity distribution only slightly varies with propagation distance. A similar transition from diffraction to localization was observed for diagonal disorder.

Typical averaged output intensity distributions are shown using a logarithmic scale in Figs. 3(b) and 3(c) for bending amplitudes below and above resonance. The distributions are triangular, indicating an exponential decay of the beam tails. Figure 3(d) illustrates how ballistic expansion in a regular array is replaced by localization with increasing disorder. The distance at which the beam width $W_\chi$ approaches its asymptotic value rapidly decreases when the level of disorder grows. At high disorder levels such distance becomes smaller than the period of the array bending $T$, so that the wave localization process mainly involves rapid spatial beatings between the excited localized Anderson modes that are nearly unaffected by the array curvature. In contrast, at lower disorder levels the value of array bending notably affects localization, whose onset is observed only after several bending periods [see Fig. 1(b)].

At the resonant bending amplitude inclusion of disorder chiefly results in suppression of beam width oscillations (Fig. 2). Notice, however, that in contrast to the case of regular arrays, where a single-site excitation exhibits complete restoration at $\xi = nT$ and the width acquires its minimum possible value [Fig. 2(a)], in disordered arrays the averaged intensity distribution extends over several array sites even at $\xi = nT$. This important effect is most visible is Fig. 2(c). Physically, it implies that the localization degree at resonance is reduced due to disorder, i.e. there is a *competition* between dynamic and Anderson localization. Indeed, at resonance in a regular array the effective coupling *vanishes* at one bending period, causing light to concentrate *nearly fully* within just one waveguide. In contrast, in the presence of disorder the effective coupling becomes *nonzero* for some waveguides because a particular value of the resonant amplitude is affected by the waveguide spacing or depth, which in turn depend on disorder. This leads to a broadening of the pattern, because the Anderson localization mechanism that counteracts diffraction is based on an interference effect and thus the corresponding Anderson modes will *always* extend over several waveguides, instead of being localized in one of them. The same effect takes place for diagonal disorder, as depicted in Fig. 2(d).

The central results of the comprehensive series of numerical experiments that we have conducted are summarized in Figures 4 and 5. On the one hand, Fig. 4 shows the evolution of the statistically averaged beam width with increasing level of off-diagonal [Fig. 4(a)] and diagonal [Fig. 4(b)] disorder, for several bending amplitudes $A<A_r$, $A=A_r$, and $A>A_r$. On the other hand, conversely, Figure 5 shows how at a fixed disorder level the localization degree of the statistically averaged beam width varies with the bending amplitude around resonance.

Figure 4 show how for non-resonant amplitudes an increase of the disorder level results in the enhancement of localization, although in the case of diagonal disorder slightly nonmonotonic dependencies $W_\chi(p_\mathrm{d})$ were obtained. Consistent with expectations, larger disorder levels are required to achieve the same degree of localization in arrays with stronger effective diffraction, i.e., larger detuning from resonance. At resonance, the presence of a small disorder causes an expansion of the averaged pattern, hence weaker localization, because the conditions for exact beam restoration are not satisfied anymore due to small fluctuations of the coupling constants between the different sites. Such expansion is slightly more noticeable for diagonal disorder [Fig. 4(b)]. In any case, at sufficiently high disorder levels, one observes transition to localization solely due to Anderson mechanism, and the width of the averaged pattern starts decreasing monotonically with increasing disorder level. Noticeable differences between curves are only observed at intermediate and small disorder levels, while at high disorder levels the statistically averaged widths for different bending amplitudes nearly coincide, thus indicating that Anderson localization strongly dominates. It should be properly appreciated that Fig. 4 summaries comprehensive series of numerical experiments conducted on a wide parameter range. Therefore, in particular, the results presented in this figure remain qualitatively similar for other sets of experimentally relevant parameters.

Figure 5 further confirms that the highest localization occurs at resonance, and shows that disorder is more detrimental for localization in the case of structures with bending amplitudes below resonance. Such a behavior of the averaged width upon variation of the bending amplitude in disordered arrays is consistent with effects caused by the strength of effective diffraction in regular arrays [Fig. 3(a)]. Namely, for a given disorder level Anderson modes become more localized when the effective coupling constant between lattice sites decreases. In the case of single-site excitation, the width of the statistically averaged output pattern is dictated by the characteristic scale of the Anderson modes and, therefore, on physical grounds it is expected to decrease when $A \to A_r$. Comparing Figs. 5(a) and 5(b) yields the conclusion that diagonal and off-diagonal disorders lead to slightly different quantitative values of the statistically averaged width, but the main result is similar in both cases, thus showing the robustness of the reported physical effect.

## 4. Conclusions

Summarizing, we explored the interplay between two fundamentally different wave-packet localization mechanisms – dynamic resonant localization induced by geometry and disorder-induced Anderson localization. We considered the particular case of optical wave propagation in periodically-curved waveguide arrays, but the results are intended to be general and hence relevant to other similar physical settings where the two mechanisms occur. We showed how the width of the statistically averaged patterns varies with the combined effects of disorder and bending, and found that disorder-induced effects may significantly enhance localization for off-resonant bending amplitudes. For suitable disorder levels, Anderson localization was found to strongly dominate over dynamic localization, a result whose importance has to be properly appreciated given the ubiquity of the involved phenomena in several areas of physics.

This work was supported by the *Severo Ochoa Excellence* program and by *Fundació Privada Cellex Barcelona*.

# Figure captions

Figure 1.  (Color online) Ensemble-averaged intensity distributions illustrating dynamics of Anderson localization in periodically curved disordered waveguide arrays with $A=9.2$ for different levels of the off-diagonal disorder $s_\mathrm{d}$. In all plots shown in the paper, the transverse and longitudinal periods are $d=2$ and $T=60$, respectively. The dynamics is shown within $\eta\in[-30.8, 49.2]$ window up to the distance $\xi=16T$. All quantities are plotted in dimensionless units.

Figure 2.  (Color online) Same as in Fig. 1, but for the bending amplitude $A=11.48$. The upper panel shows propagation in a regular array, the two middle panels show the effect of off-diagonal disorder and the lower panel corresponds to diagonal disorder. The dynamics is shown within $\eta\in[-30.8, 49.2]$ window up to the distance $\xi=16T$. All quantities are plotted in dimensionless units.

Figure 3.  (a) Width of the output beam at $\xi=960$ versus bending amplitude in a regular curved waveguide array. Circles correspond to the propagation dynamics shown in Figs. 1(a) and 2(a). Statistically averaged output intensity distributions in logarithmic scale at $\xi=960$, $s_\mathrm{d}=0.1$ for (b) $A=9.2$ and (c) $A=12.6$. (d) Statistically averaged beam width in an array with off-diagonal disorder and $A=9.2$ versus propagation distance for $s_\mathrm{d}=0$ (curve 1), $s_\mathrm{d}=0.06$ (curve 2), and $s_\mathrm{d}=0.2$ (curve 3). All quantities are plotted in dimensionless units.

Figure 4.  Statistically averaged output width at $\xi=960$ versus strength $s_\mathrm{d}$ of the off-diagonal disorder (a) and strength $p_\mathrm{d}$ of the diagonal disorder (b) for the bending amplitudes $A=9.2<A_r$ (open circles), $A=11.48=A_r$ (triangles), and $A=12.6>A_r$ (solid circles). All quantities are plotted in dimensionless units.

Figure 5.  (Color online) Statistically averaged output width versus bending amplitude (a) for off-diagonal disorder with $s_\mathrm{d}=0.05$ (rhombuses) and $s_\mathrm{d}=0.07$ (circles), and (b) for diagonal disorder with $p_\mathrm{d}=0.004$. The vertical dashed lines indicate the resonant bending amplitude $A_r=11.48$. All quantities are plotted in dimensionless units.

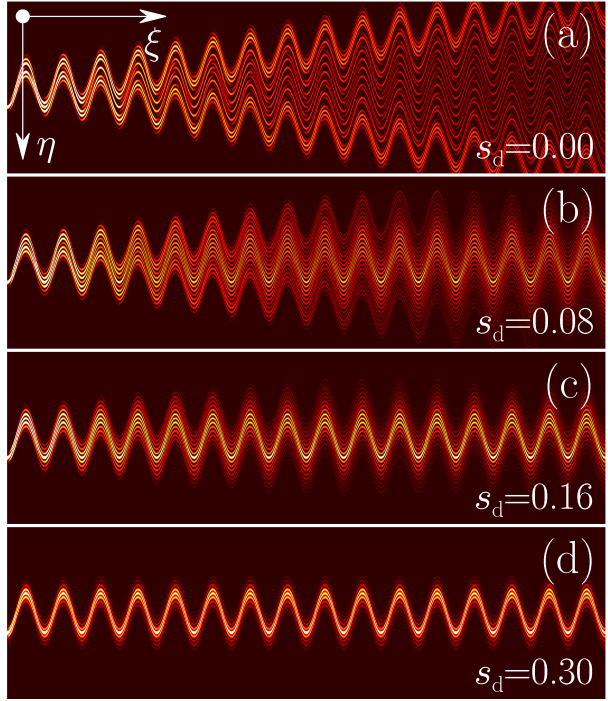

Figure 1. (Color-online) Ensemble-averaged intensity distributions illustrating dynamics of Anderson localization in periodically curved disordered waveguide arrays with $A=9.2$ for different levels of the off-diagonal disorder $s_{\rm d}$. In all plots shown in the paper, the transverse and longitudinal periods are $d=2$ and $T=60$, respectively. The dynamics is shown within $\eta \in [-30.8, 49.2]$ window up to the distance $\xi=16T$. All quantities are plotted in dimensionless units.

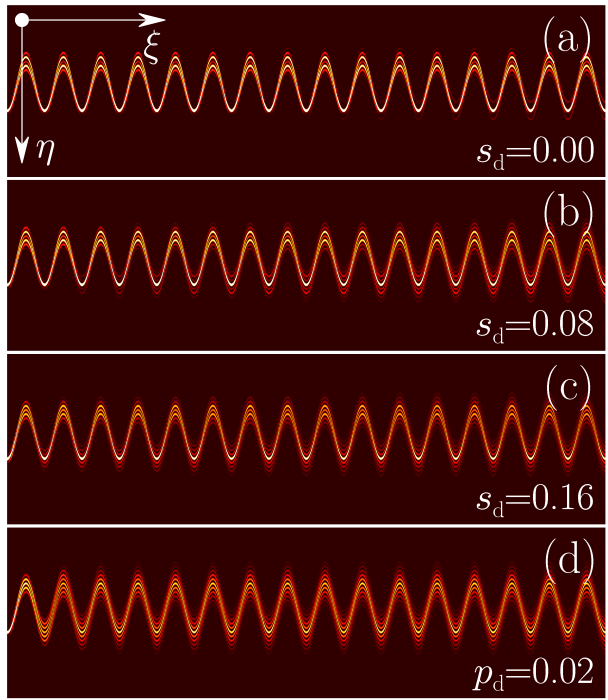

Figure 2. (Color online) Same as in Fig. 1, but for the bending amplitude $A = 11.48$. The upper panel shows propagation in a regular array, the two middle panels show the effect of off-diagonal disorder and the lower panel corresponds to diagonal disorder. The dynamics is shown within $\eta \in [-30.8, 49.2]$ window up to the distance $\xi = 16T$. All quantities are plotted in dimensionless units.

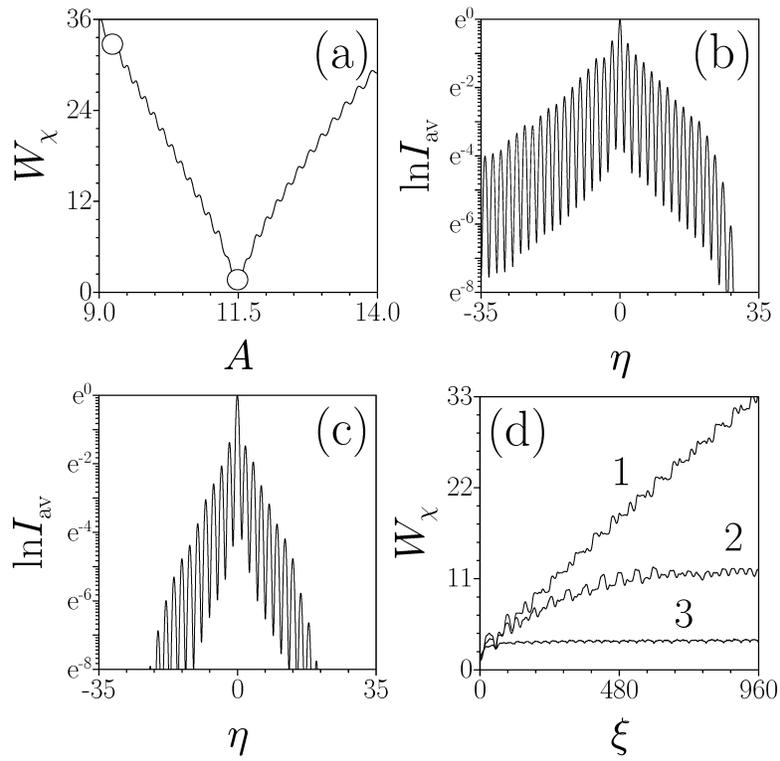

Figure 3. (a) Width of the output beam at $\xi=960$ versus bending amplitude in a regular curved waveguide array. Circles correspond to the propagation dynamics shown in Figs. 1(a) and 2(a). Statistically averaged output intensity distributions in logarithmic scale at $\xi=960$, $s_d=0.1$ for (b) $A=9.2$ and (c) $A=12.6$. (d) Statistically averaged beam width in an array with off-diagonal disorder and $A=9.2$ versus propagation distance for $s_d=0$ (curve 1), $s_d=0.06$ (curve 2), and $s_d=0.2$ (curve 3). All quantities are plotted in dimensionless units.

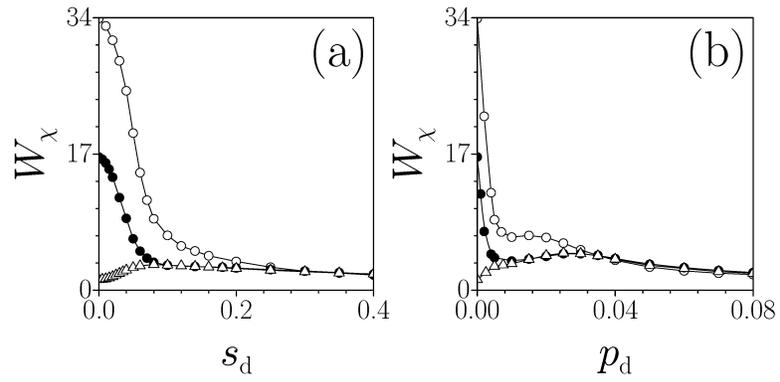

Figure 4. Statistically averaged output width at $\xi=960$ versus strength $s_d$ of the off-diagonal disorder (a) and strength $p_d$ of the diagonal disorder (b) for the bending amplitudes $A=9.2<A_r$ (open circles), $A=11.48=A_r$ (triangles), and $A=12.6>A_r$ (solid circles). All quantities are plotted in dimensionless units.

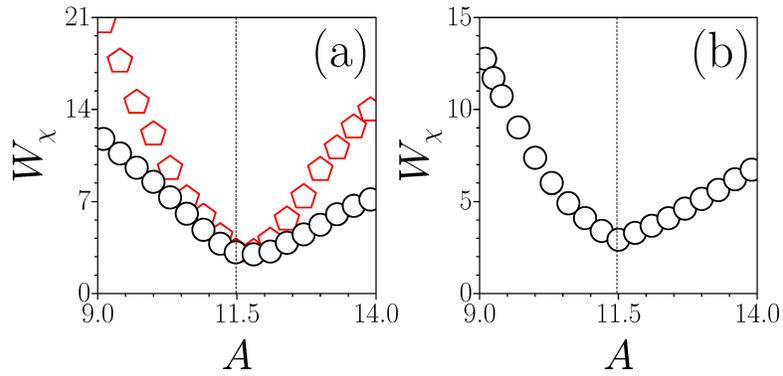

Figure 5. (Color online) Statistically averaged output width versus bending amplitude (a) for off-diagonal disorder with $s_d = 0.05$ (rhombuses) and $s_d = 0.07$ (circles), and (b) for diagonal disorder with $p_d = 0.004$. The vertical dashed lines indicate the resonant bending amplitude $A_r = 11.48$. All quantities are plotted in dimensionless units.

13